\documentclass[fleqn,twoside]{article}
\usepackage{espcrc2}
\usepackage{graphicx}
\usepackage[figuresright]{rotating}

\newcommand{\beq}{\begin{equation}}
\newcommand{\eeq}{\end{equation}}
\newcommand{\beqa}{\begin{eqnarray}}
\newcommand{\eeqa}{\end{eqnarray}}

\newcommand{\AmS}{{\protect\the\textfont2
  A\kern-.1667em\lower.5ex\hbox{M}\kern-.125emS}}
\def \osum{\,{\oplus}\,}
\def \O{{\cal O}}
\def \tr{{\rm tr}\,}
\def \etc{{\sl etc.\/}}
\def \ie{{\sl i.e.\/}}
\def \Im{{\rm Im}\,}

\title{Pressure and non-linear quark number susceptibilities in QCD}

\author{R.\ V.\ Gavai\address[DTP]{Dept. of Theoretical Physics,
        Tata Institute of Fundamental Research, Homi Bhabha Road,
        Mumbai 400005, India} and
        Sourendu Gupta\addressmark[DTP]\thanks{By whom the talk was presented}.}
       
\begin{document}

\begin{abstract}
We report the first measurements of sixth order non-linear quark
number susceptibilities in continuum QCD.  This extends our earlier
computation of the continuum limit \cite{old} of non-linear quark
number susceptibilities and the pressure in QCD for $m/T_c=0.1$.
For $T/T_c\ge1.5$, no phase transitions are observable for $\mu<30T$.
Systematics of off-diagonal susceptibilities are reported. The equation
of state in the high temperature phase of the QCD plasma is constructed.
\end{abstract}

\maketitle

The derivatives of the pressure with respect to the quark chemical potentials (one
for each flavour)---
\beq
   \chi^{(l)}_{fgh\cdots} =
     \frac{\partial^l P}{\partial\mu_f\partial\mu_g\partial\mu_h\cdots}
\label{qns}\eeq
are the quark number susceptibilities (QNS, for $l\ge2$) and the number
densities ($n$, for $l=1$). Thus, the quark number susceptibilities
measured at zero chemical potential (where direct lattice computations are
possible) allow reconstruction of the equation of state at finite chemical
potential through the usual multivariate Taylor series expansion---
\beq
   \Delta P(T,\{\mu\}) = \sum_{l;fgh\cdots} \frac1{l!}
        \chi^{(l)}_{fgh\cdots} \mu_f\mu_g\mu_h\cdots,
\label{eos}\eeq
where $\Delta P(T,\{\mu\})=P(T,\{\mu\})-P(T,\{0\})$. The coefficients for
odd $l$ vanish due to CP symmetry. The convergence of eq.\ (\ref{eos})
is spoilt by phase transitions; hence any bound on the radius of
convergence limits the location of the nearest phase transition to
the point at $\mu=0$.  Due to finite lattice spacing ambiguities in
$\chi^{(n)}_{fgh\cdots}$, and the rapid convergence of the Taylor
expansion away from a phase transition, the best way to compute the
continuum equation of state (EOS) is through the Taylor expansion
\cite{old}. We show later that, for $T>T_c$, the equation of state
\cite{fk,old,biswa} is dominated by the $l=2$ term (here $T_c$ means the
transition or crossover point for $\mu=0$). The QNS are also interesting
in themselves \cite{val}.

We report measurements of the continuum limit of the susceptibilities upto
order 6 in QCD using the quenched approximation for $T\ge1.5T_c$, where
the difference between quenched and dynamical computations is expected to
be of the order of 5--10\%. Near $T_c$, where the difference is large,
we show results obtained at finite cutoff in dynamical two-flavour
computations.

\begin{figure}[htb]
\includegraphics[scale=0.5]{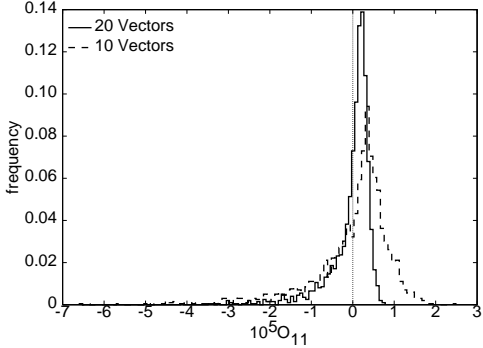}
\caption{Normalised histogram of measurements of $\O_{11}$ with different
   $N_v$ for dynamical $N_f=2$ computations at $1.05T_c$ on a $4\times10^3$
   lattice with $m/T_c=0.1$.}
\label{fg.histo}
\end{figure}

\begin{figure}[hbt]
\includegraphics[scale=0.27]{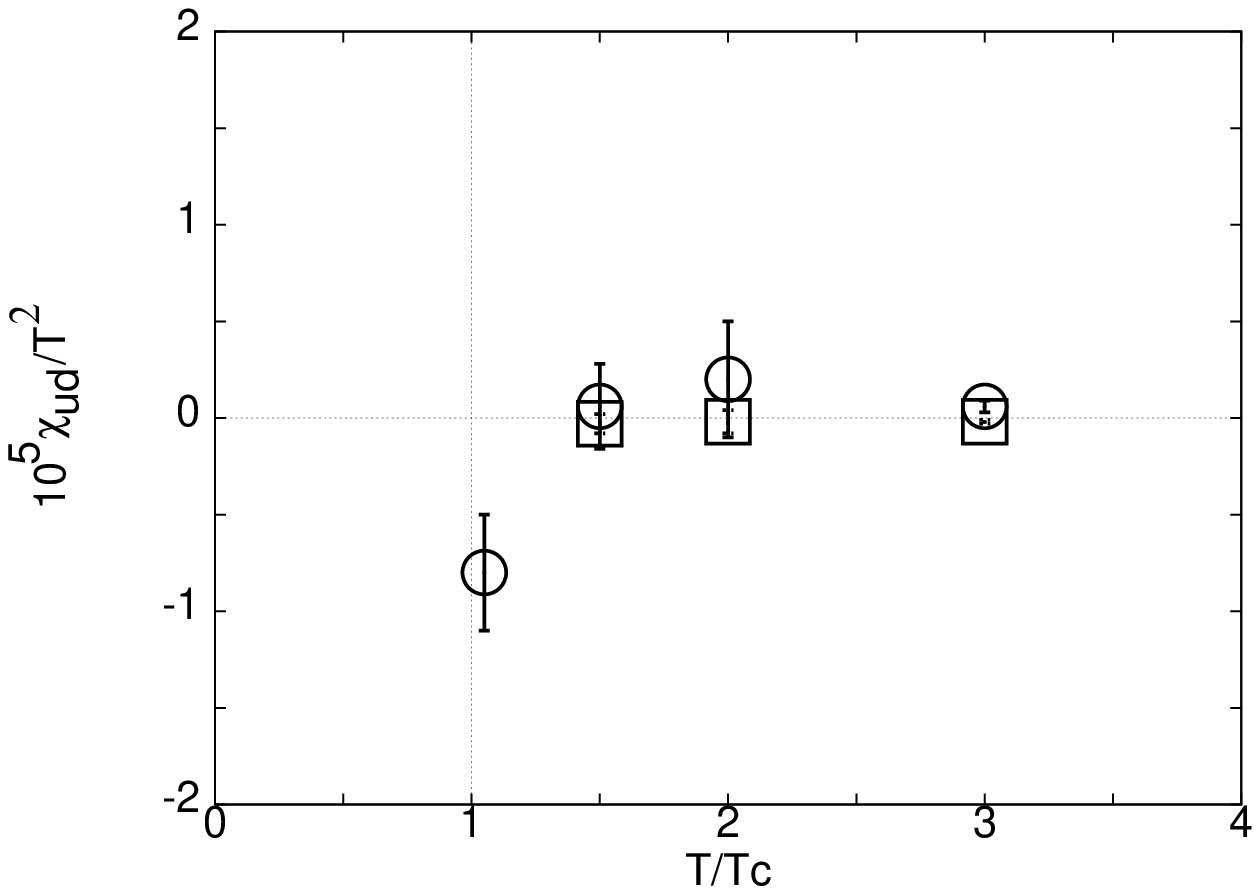}
\includegraphics[scale=0.27]{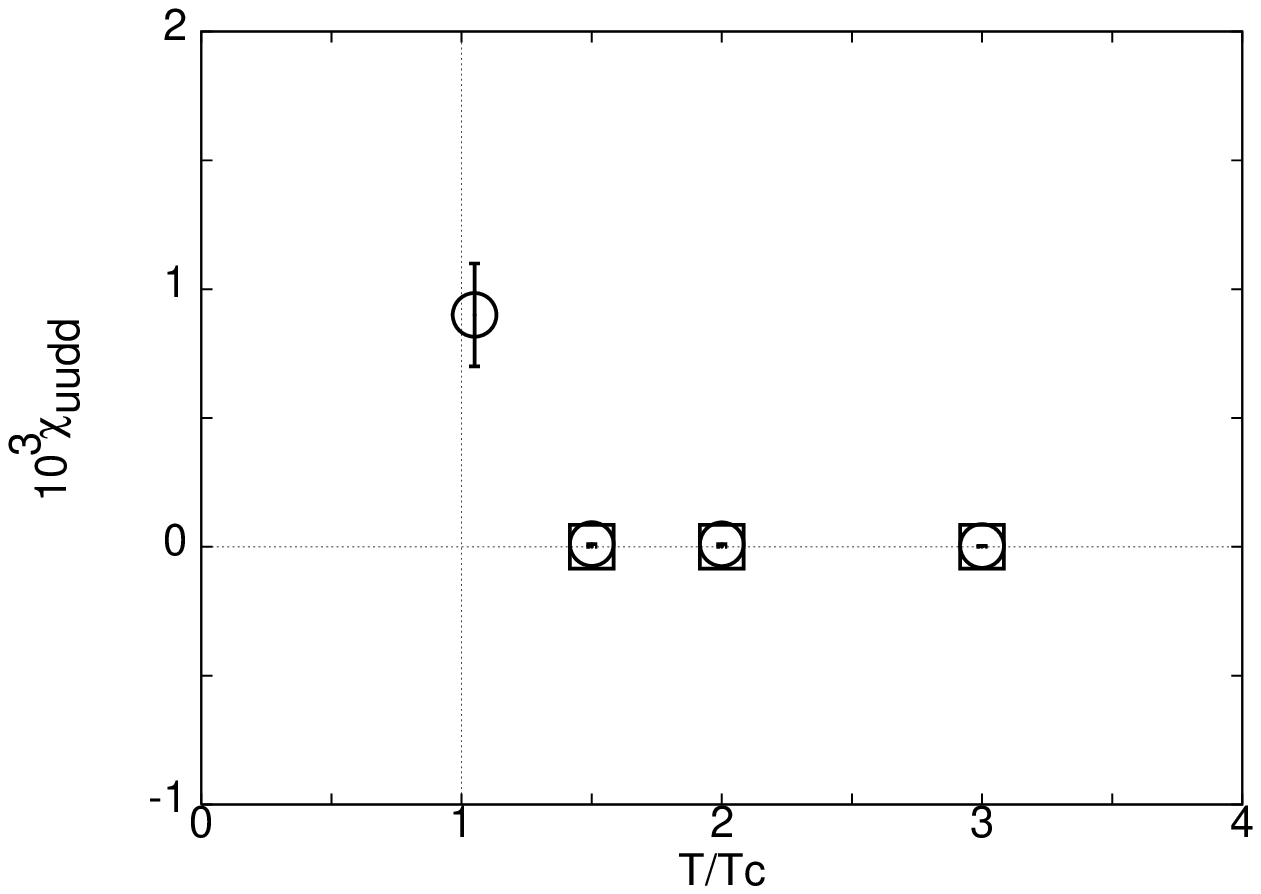}
\caption{Temperature dependence of $\chi_{ud}$ (left panel) and $\chi_{uudd}$
   (right panel). Boxes denote data for quenched QCD with $N_t=14$, and circles
   for dynamical $N_f=2$ QCD with $N_t=4$.}
\label{fg.offd}
\end{figure}

Off-diagonal susceptibilities need the computation of quantities such as
$\tr A^n$ for matrices $A$.  Independent sets of $N_v/n$ random complex
Gaussian vectors, $\{r\}$, $\{s\}$, \etc, are used for the noisy
evaluation such as $\tr A^2=\overline{\langle r|A|r\rangle\langle
s|A|s\rangle}$. For the evaluation of $\O_{11}$, where $A$ is
anti-Hermitean, the result is negative: in the $N_v\to\infty$ limit,
the histogram of noisy evaluations is skew, peaking at zero, with
a tail to the left, and vanishing on the right. As shown in Figure
\ref{fg.histo}, fairly large $N_v$ are needed to get a signal (we
used $\Im\langle r|A|r\rangle$ for noise reduction) near $T_c$. For
larger $T$ even $N_v=100$ does not give a signal \cite{old}. The
conjugate gradient stopping criterion is not a crucial parameter in this
computation.  Of the off-diagonal susceptibilities, only $\chi_{ud}$ and
$\chi_{uudd}$ were found to differ from zero significantly near $T_c$
(degenerate 3-flavour QNS such as $\chi_{uds}$ and $\chi_{udss}$ also vanish
within errors). The temperature dependence of these quantities is shown
in Figure \ref{fg.offd}. The peaks in these quantities close to $T_c$
are crucial for building up the peaks in the pressure near the critical
end point \cite{biswa}, since the diagonal susceptibilities go smoothly
to zero near $T_c$.

The 6th order diagonal QNS, $\chi_{uuuuuu}$, can be reconstructed by
standard methods from the sixth derivative of $Z$---
\beqa
\nonumber
    &&Z_{uuuuuu} = Z\langle \O_{111111} + 14\O_{11112} + 20\O_{1113}
      \biggr.\\ \nonumber &&\;\biggl.
       + 45\O_{1122} + 15\O_{114} + 60\O_{123} + 6\O_{15}
      \biggr.\\ &&\;\biggl.
       + 15\O_{222} + 15\O_{24} + 10\O_{33} + \O_6 \rangle,
\label{z6}\eeqa
where the operators other than $\O_6$ have been written out in \cite{old}.
In the notation of \cite{old} this operator is given by
\beqa
\nonumber
   &&\O_6 = -120(6\cdot1)+360(4\cdot1\osum 2)\\
\nonumber
        &&\;-210(2\cdot1\osum 2\cdot2)-60(1\osum 2\osum 1\osum 2)\\
\nonumber
        &&\;-120(3\cdot1\osum 3)+30(3\cdot2)+60(1\osum 2\osum 3)\\
\nonumber
        &&\;+60(1\osum 3\osum 2)+30(2\cdot1\osum 4)-10(2\cdot3)\\
        &&\;-15(2\osum 4)-6(1\osum 5)+(6)
\label{ord6}\eeqa
Results for the 6th order susceptibilities are shown in Table
\ref{tb.chi}.  The large finite lattice size effects shown in Table
\ref{tb.chi} and in the other QNS \cite{old} underline the need for
taking the continuum limit in order to get reliable estimates of the
location of the critical end point, where the Taylor expansion diverges
due to the growth of the higher order susceptibilities.

Estimates of the radius of convergence of eq.\ (\ref{eos}) at
the 4th and 6th orders are
\beq
   \frac{\mu_*^{(4)}}T=\sqrt{\frac{12\chi_{uu}/T^2}{\chi_{uuuu}}},
   \quad
   \frac{\mu_*^{(6)}}T=\sqrt{\frac{30\chi_{uuuu}}{T^2\chi_{uuuuuu}}}.
\label{rad}\eeq
$\mu_*^{(4)}/T\approx4.6$ in the continuum limit for $1.5\le T/T_c\le3$
\cite{old}, whereas $\mu_*^{(6)}/T\approx28$ for $N_t=14$ and is possibly
significantly larger in the continuum. For $T>T_c$, $\mu_*$
limit the phase boundary between the plasma and the appropriate
colour-superconducting phase. If the phase boundary is further off, or if,
for $T<T_c$, there is quark-hadron continuity\cite{sch}, then these two
numbers are the value of $\mu/T$ at which the two terms being compared
give equal contribution in the Taylor expansion.  At SPS energies, where
$\mu/T_c=0.45$ \cite{jc}, the 4th order term gives a 5\% correction
to $\Delta P/T^4$, and the 6th order term is totally negligible for
$T/T_c\ge1.5$. At the RHIC, where $\mu/T_c=0.15$, the leading term
contributes more than 99\% of the total.

\begin{figure}[htb]
\includegraphics[scale=0.5]{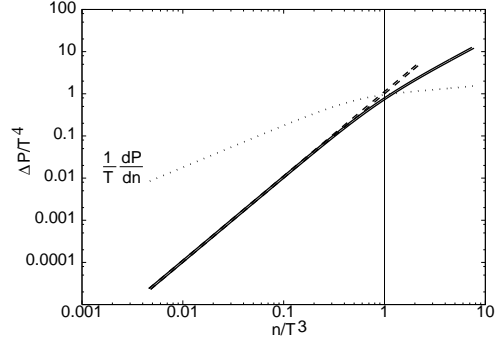}
\caption{The equation of state in the continuum limit at finite chemical potential
   for $T=2T_c$ with the leading term only (dashed line), and the first two terms
   (full line) of the Taylor series. The band shows the 1-$\sigma$ error due to
   statistical errors in the susceptibilities. The EOS at $1.5T_c$ and $3T_c$ is
   indistinguishable at this scale. The dotted line is the response function
   $\partial(P/T^4)/\partial(n/T^3)$ at fixed $\mu$ and $T$--- the instability
   in \cite{biswa} is an artifact.}
\label{fg.eos}
\end{figure}

\begin{sidewaystable}
\newcommand{\m}{\hphantom{$-$}}
\newcommand{\cc}[1]{\multicolumn{1}{c}{#1}}
\renewcommand{\arraystretch}{1.2} 
\begin{tabular*}{\textheight}{@{\extracolsep{\fill}}lllllllllllll}
\hline
QNS & $T/T_c$ & $N_t=4$ & $N_t=6$ & $N_t=8$ & $N_t=10$ & $N_t=12$ & $N_t=14$ \\
\hline
$T^2\chi_{uuuuuu}$ & 1.05 &  1.53   (9) &             &             &
                                    &             &             \\
                & 1.5 &  0.993  (3) &             &  0.142  (5) &
                         0.061  (5) &  0.03   (1) &  0.032  (8) \\
                & 2.0 &  0.931  (2) &  0.3969 (6) &  0.148  (5) &
                         0.073  (5) &  0.038  (7) &  0.02   (3) \\
                & 3.0 &  0.8683 (8) &  0.4150 (6) &  0.160  (3) &
                         0.082  (3) &  0.045  (3) &  0.029  (8) \\
\hline
$T^2\chi_{uuuudd}$ & 1.05 & 0.0148 (7) &               &               &
                                      &               &               \\
                & 1.5 & 0.0364 (3)    &               & 0.0208 (4)    &
                        0.0106 (2)    & 0.00653 (9)   & 0.00435 (5)   \\
                & 2.0 & 0.0385 (2)    & 0.0684 (6)    & 0.0228 (1)    &
                        0.0116 (1)    & 0.00692 (4)   & 0.00482 (7)   \\
                & 3.0 & 0.0406 (2)    & 0.0772 (5)    & 0.0255 (5)    &
                        0.0120 (1)    & 0.0075 (1)    & 0.00498 (7)   \\
\hline
$T^2\chi_{uuddss}/10^{-6}$ & 1.05 & $-108 (10)$ &               &               &
                                      &               &               \\
                & 1.5 &  $-459 (8)$ &               & $-9.06 (9)$ &
                         $-1.83 (5)$ & $-0.54 (1)$ & $-0.188 (4)$ \\
                & 2.0 & $-510 (4)$ & $-110 (2)$ & $-10.1 (2)$ &
                         $-2.08 (5)$ & $-0.60 (1)$ &  $-0.214 (8)$ \\
                & 3.0 &  $-565 (6)$ & $-128 (2)6$ &  $-11.4 (8)$ &
                        $-2.10 (5)$ & $-0.65 (2)$ &  $-0.22 (6)$ \\
\hline
\end{tabular*}\\[2pt]
\caption{Various sixth order susceptibilities evaluated on the configurations
   used in \cite{old} for quark masses $m/T_c=0.1$. Results at $T/T_c=1.05$
   are for $N_f=2$ dynamical QCD, the remainder for quenched QCD. The continuum
   limit of $\chi_{uuuuuu}$ is consistent with zero at the 99\% CL.}
\label{tb.chi}
\end{sidewaystable}

The relation between $\Delta P(T,\{\mu\})/T^4$ and the quark
number density $n(T,\{\mu\})/T^3$ is the equation of state
\cite{biswa}. Since $n$ is the first derivative of the pressure, the
EOS is completely determined once the susceptibilities are known. As
shown in Figure \ref{fg.eos}, the difference between the leading term
involving $\chi_{uu}$, and the next term, becomes important only when
$n/T^3\approx1$, \ie, for $\mu\approx T$. The remaining terms change the
predictions by less than the statistical error band shown.  Away from
$T_c$, this justifies, {\sl post-facto\/}, the scaling of the EOS obtained
by reweighting at $N_t=4$ to give the continuum limit, as performed in
\cite{fk}. Finally, Figure \ref{fg.eos} shows that the measurement of
susceptibilities is the simplest and most accurate route to the EOS in
this range of $T$ and $\mu$ which is important for experiments.

\end{document}